\documentclass[twocolumn,showpacs,floatfix,preprintnumbers,amsmath,amssymb,prl,superscriptaddress]{revtex4}
\usepackage{graphicx}
\usepackage{amssymb}
\usepackage{multirow}
\usepackage{epstopdf}
\usepackage{color}

\begin{document}

\title{The fundamental differences between Quantum Spin Hall edge-states at zig-zag and armchair terminations of honeycomb and ruby nets}
\author{Laura Cano-Cort\'es}
\affiliation{Institute for Theoretical Solid State Physics, IFW Dresden, 01171 Dresden, Germany}
\author{Carmine Ortix}
\affiliation{Institute for Theoretical Solid State Physics, IFW Dresden, 01171 Dresden, Germany}
\author{Jeroen van den Brink}
\affiliation{Institute for Theoretical Solid State Physics, IFW Dresden, 01171 Dresden, Germany}
\affiliation{Department of Physics, TU Dresden, D-01062 Dresden, Germany}

\date{\today}

\begin{abstract}
Combining an analytical and numerical approach we investigate the dispersion of the topologically protected spin-filtered edge-states of the Quantum Spin Hall state on honeycomb and ruby nets with zigzag (ZZ) and armchair (AC) edges. We show that the Fermi velocity of the helical edge-states on ZZ edges increases linearly with the strength of the spin-orbit coupling (SOC) whereas for AC edges the Fermi velocity is {\it  independent} of the SOC. 
Also the decay length of edge states into the bulk is dramatically different for AC and ZZ edges, displaying an inverse functional dependence on the SOC. 
\end{abstract}
\maketitle

{\it Introduction} -- In their seminal paper \cite{Kane05a},  Kane and  Mele have shown that spin orbit coupling (SOC) in a  single plane of graphene leads to a time-reversal invariant Quantum Spin Hall (QSH) state that is characterized by a bulk energy gap and a pair of topologically protected gapless edge-states. However, the SOC energy scale in graphene is so tiny that the predicted gap \cite{Huertas06} is merely $\sim0.01 K$. It is therefore not possible in practice to establish the existence of the underlying ${\cal Z}_2$ topological order in graphene \cite{Kane05b,Fu07,Hasan10}
so that the hunt for the QSH effect was continued in other materials. The theoretical prediction \cite{Bernevig06} and experimental observation \cite{Honig07} of the QSH effect in HgTe thin films have firmly categorized this material as the sought-after two-dimensional (2D) ${\cal Z}_2$ topological insulator (TI). Having a zinc-blende crystal structure this material is of course very different from graphene from both an electronic and a structural point of view.

\begin{figure}
\includegraphics[clip,width=.9\columnwidth,keepaspectratio]{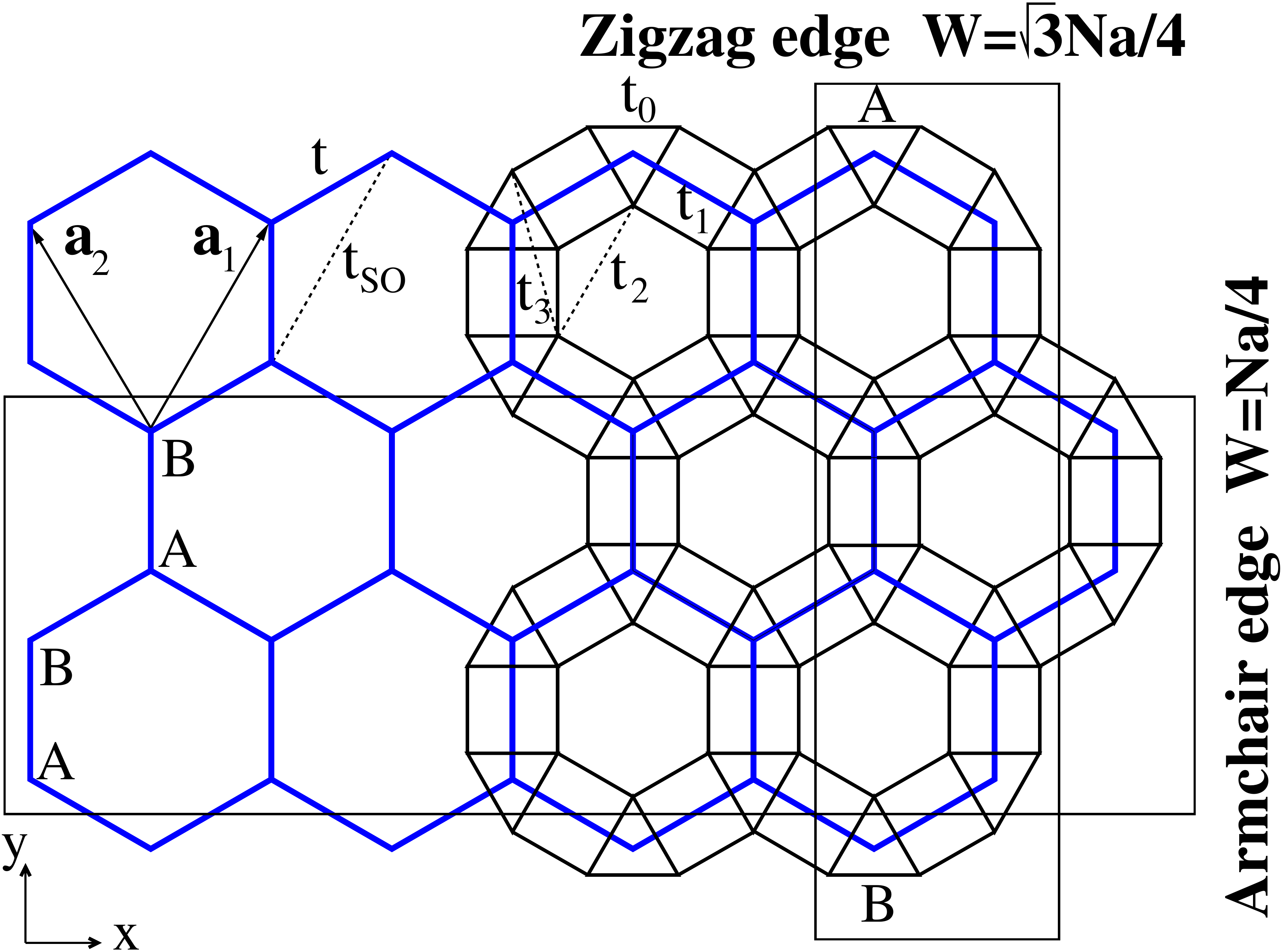}
\caption{(Color online) Lattice structure of honeycomb (blue) and ruby nets (black). The primitive lattice vectors are ${\bf a}_1$ and ${\bf a}_2$. The nearest-neighbor hopping is $t$ ($t_0$, $t_1$) and the SOC strength $t_{SO}$ ($t_2$, $t_3$) for the honeycomb (ruby) lattice. 
Top and bottom are zigzag (ZZ) edges and left and right are armchair (AC) edges. The width \cite{footnote1} of the ribbons as a function of the number
of atoms in the unit cell $N$ is indicated by $W$. 
}
\label{lattice}
\end{figure}

The recently discovered topological insulator Bi$_{14}$Rh$_{3}$I$_9$, however, provides an entirely novel platform for the observation of the QSH effect in graphene-like systems with a honeycomb structure \cite{Rasche13}. Bi$_{14}$Rh$_{3}$I$_9$ consists of stacks of bismuth-based layers each of which forms a honeycomb net composed of RhBi$_{8}$ cubes. These cubes form what is commonly referred to as a ruby lattice, see Fig.~\ref{lattice}, which has the same point group symmetry as the hexagonal graphene honeycomb lattice. Each such a layer of Bi$_{14}$Rh$_{3}$I$_9$ forms a 2D ${\cal Z}_2$ TI, with a large spin-orbit gap of $\sim2400 K$ due to the strong bismuth-related SOC \cite{Rasche13}. The gap being six order of magnitudes larger than  graphene opens the avenue for the actual observation of the QSH effect in a hexagonal graphene-like system. 

For a future use of this QSH effect a fundamental understanding of the topologically-protected spin-polarized edge-states is essential. We have therefore investigated the electronic characteristics of these topological edge-states in both honeycomb and ruby lattices in the presence of SOC. 
We find in these hexagonal systems a dramatic dependence of the edge-state dispersion, decay length and Fermi velocity on the edge geometry. While for a zigzag (ZZ) termination the Fermi velocity of the edge-states critically depends on the size of the spin-orbit gap, armchair (AC) edge-states exhibit a linear dispersion with a velocity that is independent of strength of the SOC. We show that indeed in simple honeycomb nets the Fermi velocity of AC-edges corresponds exactly to the Fermi velocity of the bulk massless Dirac fermions in absence of SOC. Surprisingly, we find that the edge-state decay lengths at ZZ- and AC-edges depend on the SOC strength in an opposite manner: while the first one grows with the SOC, the other shrinks.
This emphasizes the fundamentally different electronic properties of Quantum Spin Hall edge-states at zig-zag and armchair terminations of hexagonal lattices.

{\it Model} -- We start from the well-known tight-binding Hamiltonian for graphene that includes the effect of the SOC via a next-nearest-neighbor hopping \cite{Kane05a},
\begin{eqnarray}
  \label{eq:HTB}
  H &=&  - t \sum_{\langle ij \rangle \sigma} c_{i\sigma}^{\dagger} c_{j\sigma} 
  + it_{SO} \sum_{\langle\langle ij \rangle\rangle \alpha \beta} \nu_{ij} s_{\alpha \beta}^{z} c_{i\alpha}^{\dagger} c_{j\beta},
\end{eqnarray}
where $c_{i\sigma}^{\dagger}$ and $c_{i\sigma}$ are, respectively, creation and annihilation operators of an electron on site $i$ with spin $\sigma$.
The first term corresponds to nearest neighbor hopping interaction, whereas the second term connects second neighbors with a spin dependent amplitude.
The quantity $\nu_{ij}$=$+1(-1)$ if the electrons makes a left (right) turn on the lattice 
during the second-neighbor hopping.  
As indicated in Fig.~\ref{lattice}, $t$ is the nearest-neighbor hopping parameter and $t_{SO}$ is the second nearest-neighbor parameter 
within the two-dimensional honeycomb sheet, and parametrizes the strength of the SOC.

In the Bi-Rh sheets of Bi$_{14}$Rh$_{3}$I$_9$ the Bi atoms form a two-dimensional ruby lattice that can be thought of as a decorated honeycomb net as shown in Fig.~\ref{lattice}. The resulting ruby lattice, as the honeycomb one, belongs to the group of 11 Archimedean lattices, which represent the prototypes of two-dimensional arrangements of regular polygons \cite{Suding99}. It has a geometric unit cell with 6-sites and an underlying triangular Bravais lattice containing two non-equivalent nearest-neighbor bonds.
The first Brillouin zone is analogue to the one for the honeycomb lattice, with high symmetry points $\Gamma$, $M$ and $K$.
The nearest-neighbor hopping parameters are denoted by $t_0$ and $t_1$, and the second nearest-neighbor parameters by $t_2$ and $t_3$. For simplicity we consider the symmetric case $t_0$=$t_1$=$t$ and $t_2$=$t_3$=$t_{SO}$.  
The real-space triangular Bravais vectors of both lattices are  ${\bf a}_1$=$a(1/2,\sqrt{3}/2)$, 
${\bf a}_2$=$a(-1/2,\sqrt{3}/2)$,
and the reciprocal lattice basis vectors are  
${\bf b}_1$=$ 2\pi (1 , 1/ \sqrt{3}) / a$ and 
${\bf b}_2$=$  2 \pi (-1, 1 / \sqrt{3}) / a $. 

\begin{figure}
\includegraphics[clip,width=\columnwidth,keepaspectratio]{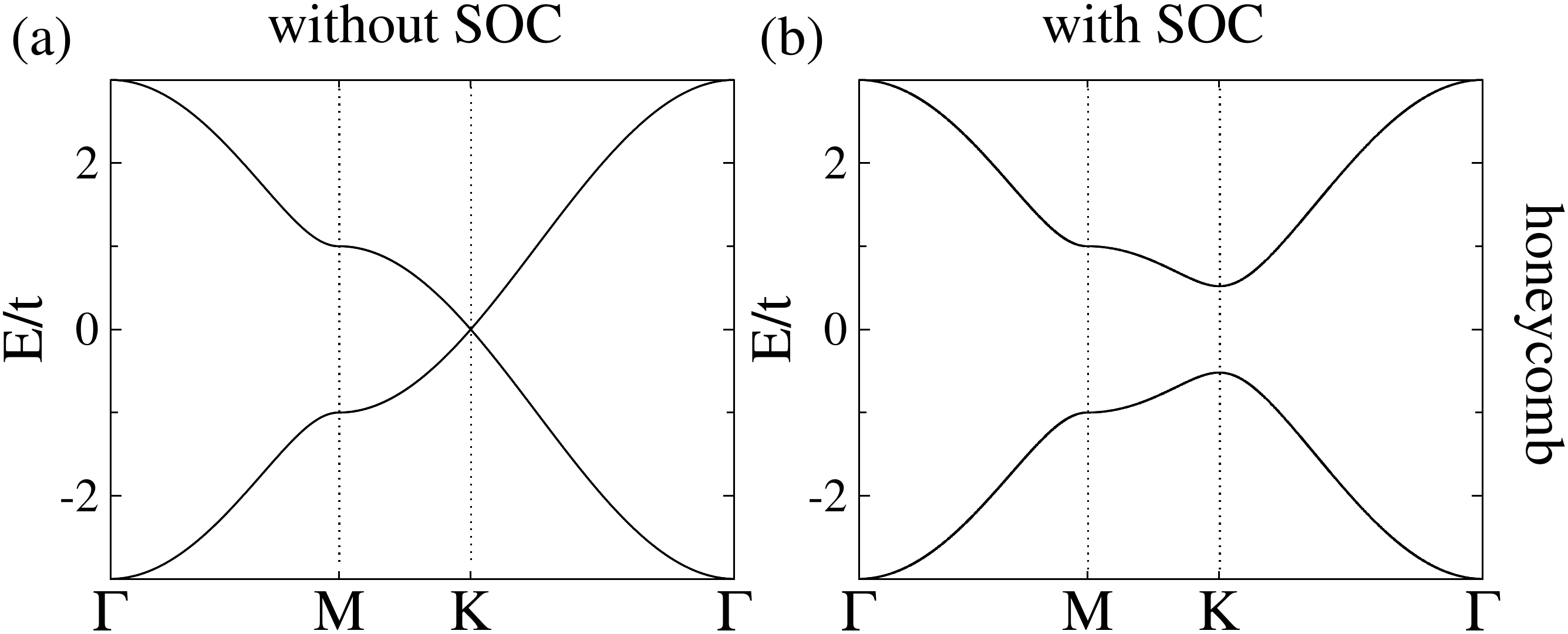}
\includegraphics[clip,width=\columnwidth,keepaspectratio]{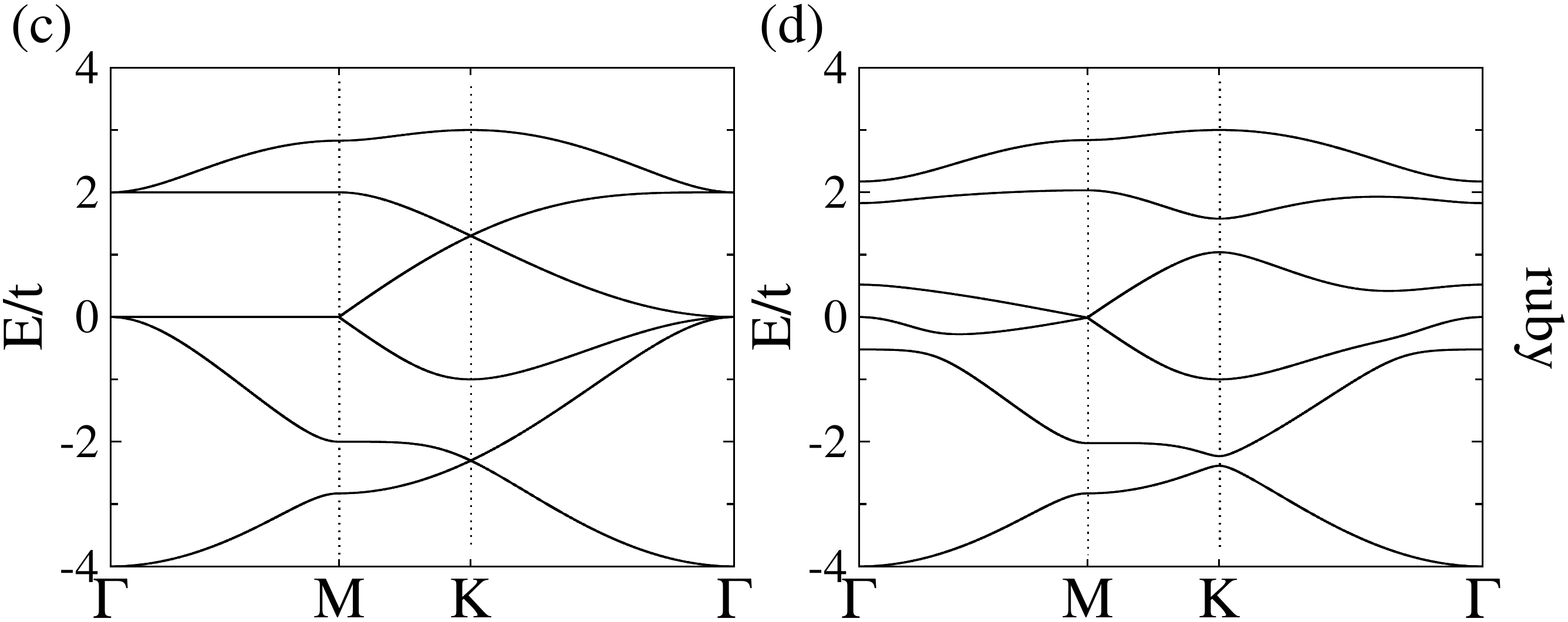}
\caption{Bulk energy bands for the honeycomb and the ruby lattices without (a), (c) and  with (b), (d) SOC.  
 Energies are measured in units of the  nearest-neightbor hopping $t$.  SOC is introduced via $t_{SO}=0.1t$.
Note that for both lattices, there is a Dirac point at $K$ in absence of SOC, opening a gap once 
this hopping is switched on. This gap opens at different fillings of the bands for the ruby lattice (1/6 and 4/6) and 
at 1/2 filling for the honeycomb case.} 
\label{bulk}
\end{figure}

Diagonalizing the Hamiltonian \eqref{eq:HTB} results in the bulk energy bands for the honeycomb and ruby nets as shown in Fig.~\ref{bulk}.
The ${\cal C}_{6v}$ point group symmetry shared by both lattices leads to the presence of massless Dirac fermions at the inequivalent ${\bf K}$ and ${\bf K}^{\prime}$ points of the Brillouin zone when the strength of the SOC vanishes.
For the ruby lattice, the Dirac points at the ${\bf K}$ (${\bf K}^{\prime}$) point appears only for 1/6 and 4/6 filling of the bands whereas a  quadratic band touching point at $\Gamma$ for 3/6 and 5/6 fillings is observed. 
The degeneracy at the Dirac points is lifted by the SOC driving the system into a topologically non-trivial QSH phase. 
For the honeycomb lattice   the presence of topological order  does not depend on the SOC strength \cite{Fu07} -- the role of the SOC is to ensure a finite gap everywhere in the Brillouin zone.
On the contrary, for honeycomb ruby nets a direct computation of the ${\cal Z}_2$ topological invariant shows that a topologically non-trivial QSH phase is stabilized for definite sets of  tight-binding parameters. For the above-mentioned symmetric choice the QSH phase occurs both at $4/6$ and $1/6$ fillings for  $0< t_{SO}/ t <0.29$:  at  $t_{SO} /t =0.29$ one finds a topological phase transition to a trivial band insulator at $4/6$ filling \cite{Hu11}.

\begin{figure}
\includegraphics[clip,width=\columnwidth,keepaspectratio]{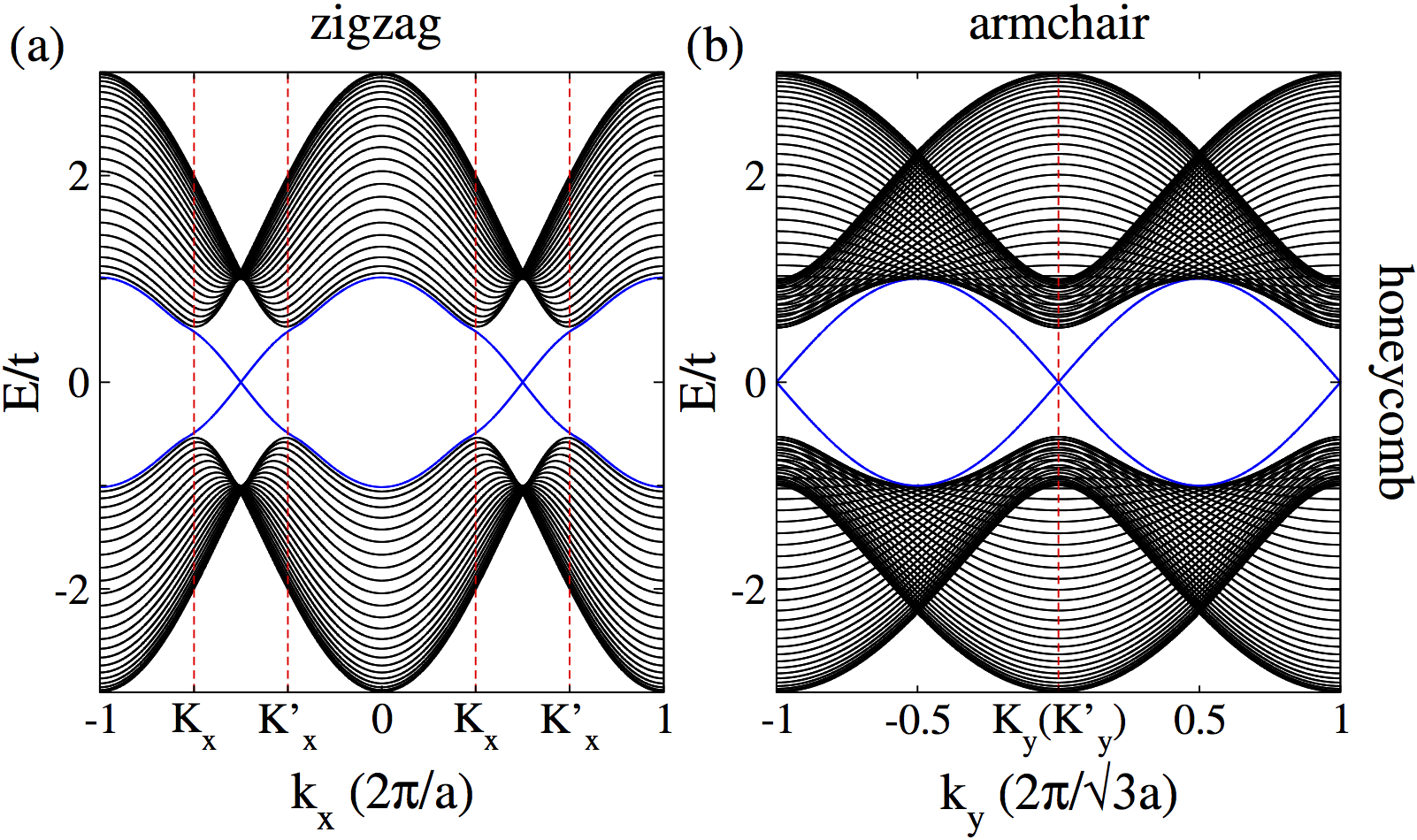}
\includegraphics[clip,width=\columnwidth,keepaspectratio]{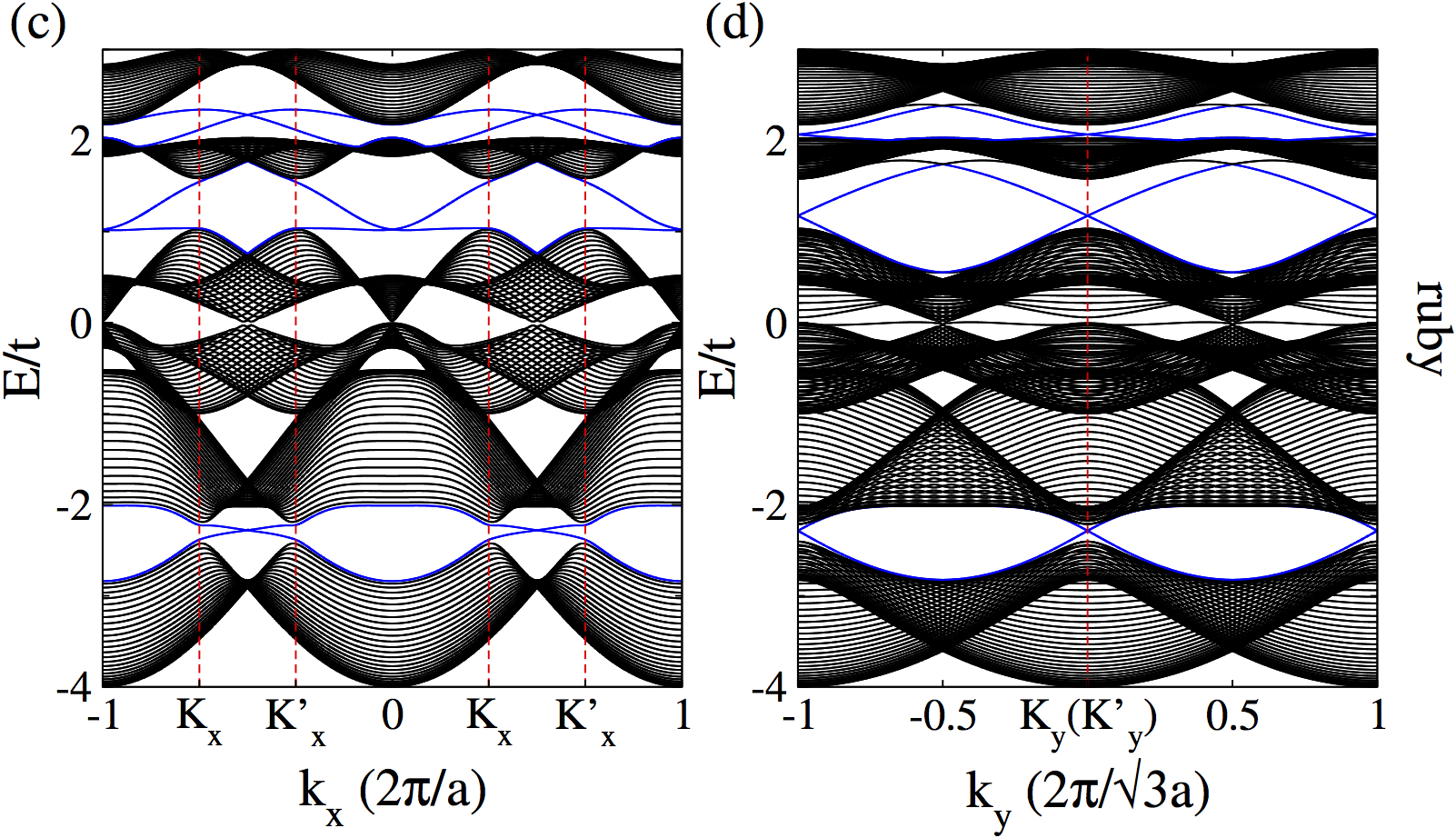}
\caption{(Color online) Energy bands for honeycomb and ruby ZZ  (a),(c), and AC
(b),(d) terminated ribbons in presence of SOC . Energies are measured in units of the 
nearest-neightbor hopping amplitude, $t$, and the SOC effects are introduced via  $t_{SO}=0.1t$.
The blue bands correspond to the edge-states crossing the gap, at 1/2 filling for honeycomb and 
for the ruby lattice at 1/6, 4/6 and 5/6 fillings. 
}
\label{edges}
\end{figure}

In order to characterize the properties of the topologically-protected edge-states in the QSH phase on honeycomb and  ruby nets we consider ribbons of the two lattices with different terminations as shown in Fig.~\ref{lattice}. In perfect analogy with the honeycomb lattice  the ruby lattice also exhibits ZZ  and AC edge terminations. The ensuing dispersions of the ribbons are summarized in Fig.~\ref{edges}.   
For honeycomb ribbons edge states appear inside the bulk spin-orbit gap at half-filling.
For the ruby lattice, we observe edge-states for  1/6, 4/6 and 5/6 fillings in agreement with the calculated ${\cal Z}_2$ topological invariant.   

{\it Fermi velocity at ZZ edges} -- The geometry of ZZ nanoribbons is shown in Fig.~\ref{lattice}, where the edge runs along $x$-axis, so that the system is translational invariant along this direction. The corresponding wavefunctions exist in the space $0<y<W$. In the absence of  SOC, the band structure of the zigzag terminated ribbons exhibits four-fold degenerate (2 spins times 2 edges) edge-localized states at zero energy in the honeycomb lattice \cite{Brey06,Akhmerov08}, appearing in perfect analogy at 4/6 and 1/6 fillings in the  ruby net [see the Supplemental Material] where close to the Fermi energy a Dirac point in the bulk band structure occurs.
These edge states connect momenta of the 1D Brillouin zone corresponding precisely to the projections of the ${\bf K}$ and ${\bf K}^{\prime}$ points of the bulk 2D Brillouin zone. 
Indeed we find that these states connect the two momenta $K_x$=$ 2\pi / (3a) $ and  $K^{\prime}_x$=$ -2\pi /(3a) $  to the edge of the 1D Brillouin zone for honeycomb  terminated ribbons as well as ruby ribbons at 1/6 filling -- in which case the edge states form almost flat bands -- whereas at 4/6 filling more dispersive edge states of the ruby ribbons interconnect $K_x$ to $K^{\prime}_x$. 

\begin{figure}
\includegraphics[clip,width=\columnwidth,keepaspectratio]{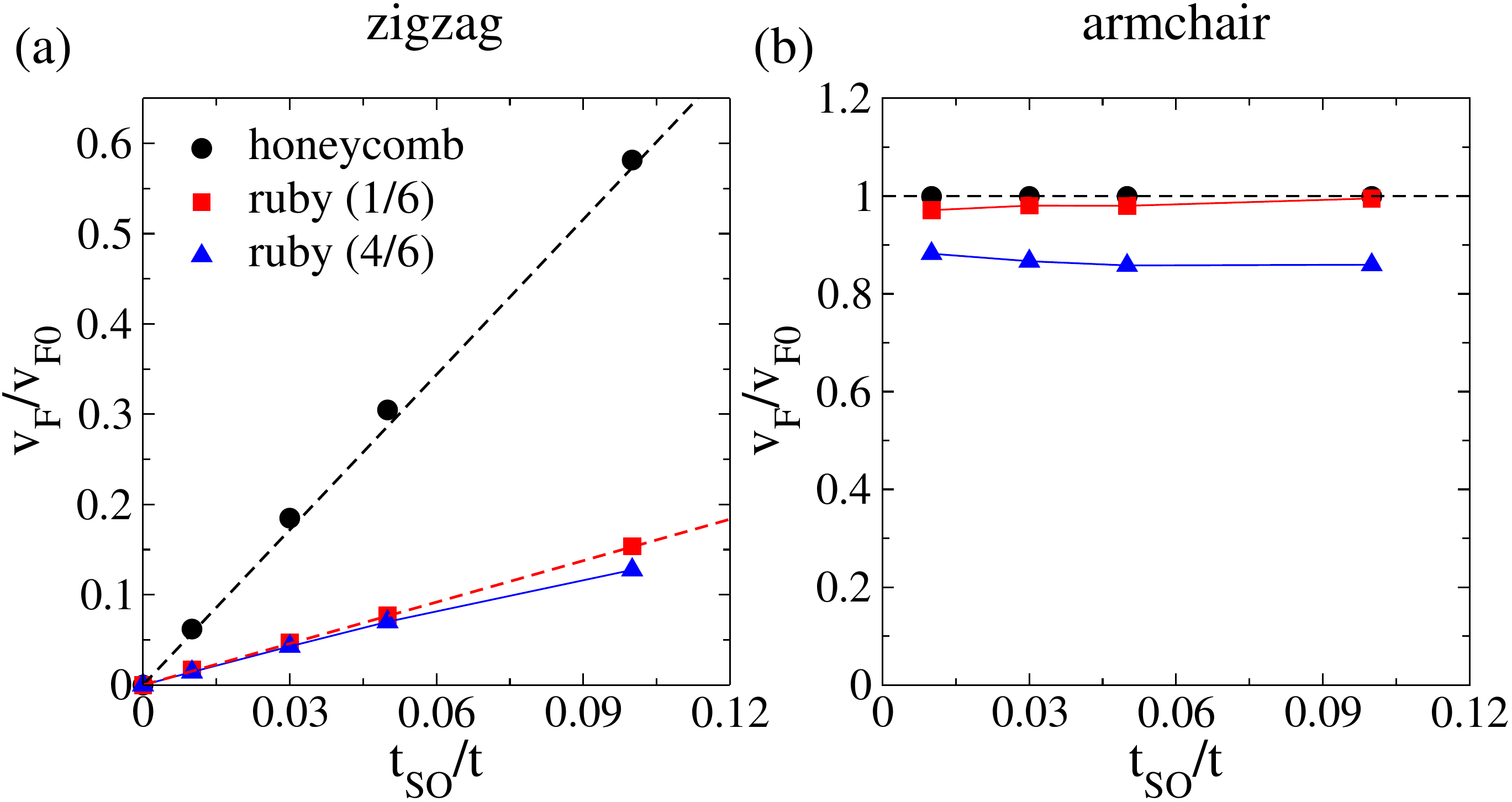}
\includegraphics[clip,width=\columnwidth,keepaspectratio]{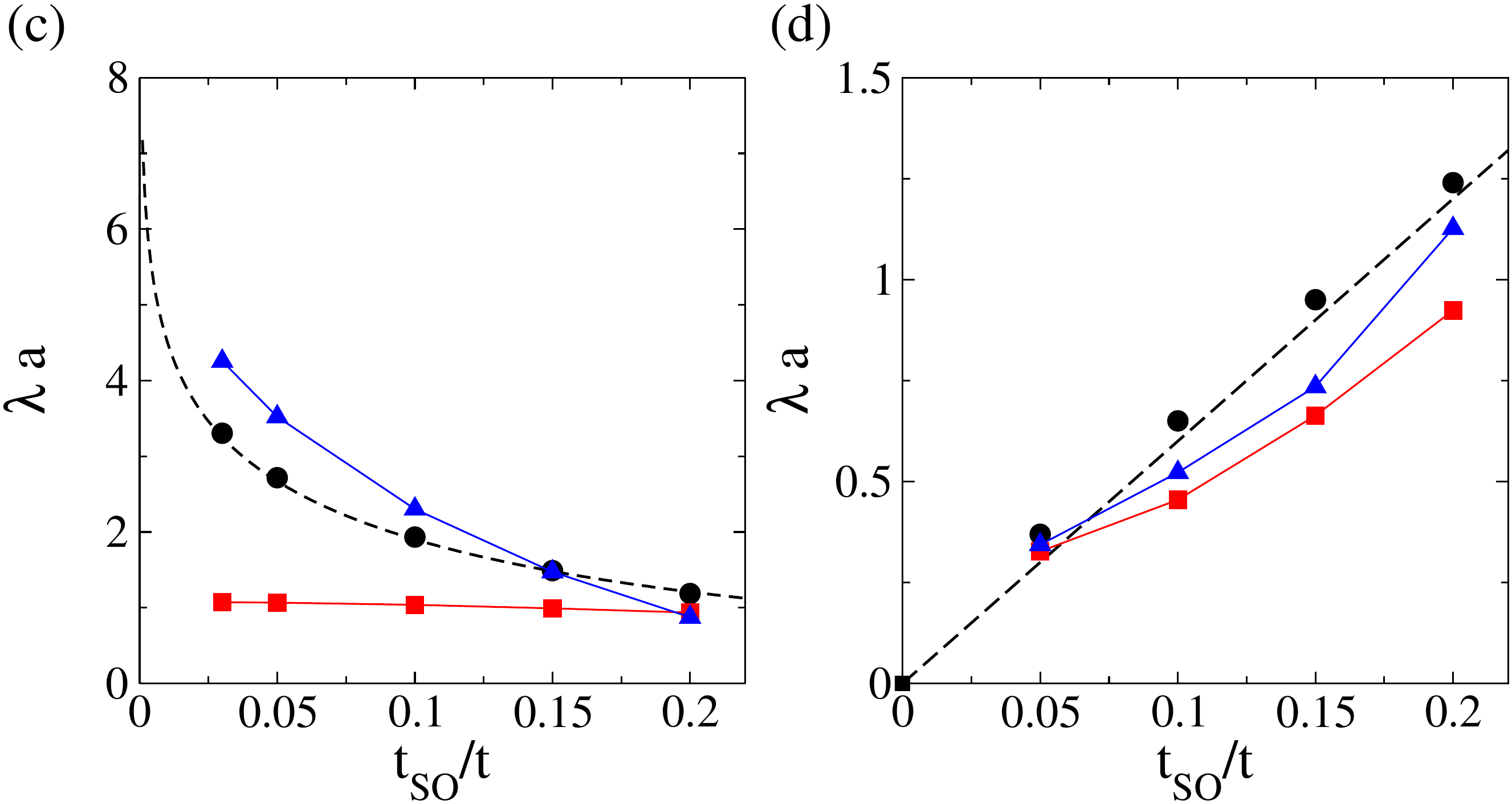}
\caption{(Color online) (a),(b): Fermi velocity of the edge states in units of the Fermi velocity of the bulk Dirac fermions
as a function of the
spin-orbit coupling, $t_{SO}$, for honeycomb and ruby zigzag (a) and armchair (b) ribbons.
(c), (d): Inverse decay length $\lambda$
for  zigzag (c) and armchair (d) ribbons as a function of the SOC.
Depending on the edge geometry $\lambda$ decreases (c) or increases (d) with $t_{SO}$. 
The dashed lines represent the analytical results.} 
\label{vfermi}
\end{figure}

When the SOC is introduced via  $t_{SO}$ helical edge states  lying in the bulk spin-orbit gap appear. At the ZZ edges their corresponding Fermi velocity increases linearly with the strength of the SOC. 
This is explicitly shown in Fig.~\ref{vfermi}(a) where we plot  the Fermi velocity  $v_F$ of the honeycomb and ruby ZZ edge-states as a function of the SOC $t_{SO}$.
For honeycomb and ruby ZZ terminated ribbons at 1/6 filling the linear dependence of $v_F$ on $t_{SO}$ can be estimated by the ratio of the spin-orbit gap and the distance between
$K_x$ and $K'_x$ points \cite{Autes12} (see Fig.~\ref{edges}).  
Therefore, $v_F= \Delta_{SO} / \Delta_{KK'}$, where $\Delta_{KK'}= 2\pi/ (3a)$. Calculating the velocity as a function of the bare 
Fermi velocity $v_{F 0 }$ of the massless bulk Dirac fermions in the absence of SOC while taking into account the dependence of the gap $\Delta_{SO}$ on 
$t_{SO}$,  we obtain
\begin{equation}
  \dfrac {v_{F}}{v_{F0}}= \alpha_{H,R} \dfrac {t_{SO}}{t}
\end{equation}
where for the honeycomb lattice $\alpha_{H} = 18 / \pi$ as it follows immediately considering that $v_{F 0 } = \sqrt{3} t a / 2$ and $\Delta_{SO} = 6 \sqrt{3} t_{SO}$, whereas for the ZZ edge states of ruby ribbons at 1/6 filling $\alpha_{R} = 4.808/\pi$.  These analytical results [c.f. the dashed lines in Fig.~\ref{vfermi}(a)] are in excellent agreement with the numerical ones. 
The edge states of ZZ terminated ruby ribbons at 4/6 filling show instead a more complicated functional dependence on the momentum:  we observe that terms up to $k^3$ are comparable in magnitude with the Fermi velocity even if the resulting $v_F$ is remarkably  close to the Fermi velocity of the 1/6 filling edge states, see Fig.~\ref{vfermi}(a). 

{\it Fermi velocity at AC edges} -- Having established the linear dependence of the Fermi velocity of the edge-states in ZZ terminated ribbons on the SOC, we now turn to discuss the properties of the edge-states in AC terminated ribbons. In the case of an AC termination, as shown in Fig.~\ref{lattice}, the edge is parallel to $y$-axis, and the 
wavefunctions exist in the space $0<x<W$. The two Dirac points are projected 
onto the time-reversal invariant point $k_y=0$ of the 1D Brillouin zone where the edge states form a Kramer's doublet. As a result, their dispersion away but close to $k_y=0$ can be analysed in the  ${\bf k.p}$ approximation \cite{DiVincenzo84,Ando05,Brey06} once SOC is explicitly taken into account.

To this end,  
we consider the effective Dirac equation  \cite{Kane05a} for the states near the ${\bf K}$ and ${\bf K}'$ points of the hexagonal Brillouin zone of the honeycomb lattice
\begin{equation}
{\cal H} =   v_{F 0} \left(\tau_z \sigma_x k_x +  \sigma_y k_y \right) + 3 \sqrt{3} t_{SO}  \sigma_z \tau_z s_z, 
\end{equation}
acting on a two-component spinorial wavefunction. In the equation above, the $\sigma$'s  are Pauli matrices with $\tau_z = \pm $ for states at the ${\bf K}$ (${\bf K}'$) points, $s_z$ representing the electron's spin and $\sigma_z = \pm$ describing states on the $A$ ($B$) sublattice (see Fig.~\ref{lattice}). 
To find the wave function for the ribbon with AC  edges we replace $k_x \rightarrow -i \partial_x$ in the Hamiltonian above \cite{Brey06}. The general solution is found by using the ansatz for the spinorial wave function $\psi^{s_z } _{\tau_z }(x)= \chi_{\tau_z}^{s_z} e^{\lambda x}$.
A real value of $\lambda$ yields evanescent waves corresponding to the edge-states. 
The wavefunctions $\psi$ have to meet the boundary condition at the armchair edges. For the honeycomb lattice, the correct boundary condition 
can be found by considering that the armchair termination consists of a line of $A-B$ dimers at $x=0$ and $x=W + a /2$. 
To do this, we admix valleys \cite{Brey06} and require $\psi^{s_z}_{+}(x=0) \equiv  \psi^{s_z}_{-}(x=0)$ and $\psi^{s_z}_{+}(x=W+ a/2) \equiv \psi^{s_z}_{-}(x=W+ a/2) \mathrm{e}^{i \Delta K \left(W + a/2 \right)}$, where $\Delta K = 4 \pi / ( 3 a)$. For ribbons whose width $(W + a/2) \Delta K = 2 \pi n$ with $n$ integer --  a condition which, in the absence of SOC, leads to a zero energy mode \cite{Brey06} --  we obtain two solutions for the energy near the ${\bf K}$(${\bf K}'$) point: $\epsilon_1 = \pm k_y v_F$ and $\epsilon_2 = \pm \sqrt {k^2_y v^2_F+27t^2_{SO}}$. The first one corresponds to edge states  lying in the bulk spin-orbit gap in which case $\lambda=3 \sqrt{3} t_{SO} / v_F$, and the second energy is the one describing the bulk bands as in this case $\lambda=0$.
As a result, the Fermi velocity of the AC edge-states is independent on the strength of the SOC and corresponds to the bare Fermi velocity of the Dirac fermions, in perfect analogy with semi-infinite AC-terminated ribbons \cite{Prada11,Zarea07}. As shown in Fig.~\ref{vfermi} we find a perfect agreement with the tight-binding calculations on honeycomb armchair-terminated ribbons and, quite remarkably that  for the ruby lattice $v_F$ is also independent of $t_{SO}$  as well. 

{\it Decay length of edge states} -- The dependence of the Fermi velocity on the strength of the SOC being completely different for ZZ and AC edges, raises the question how different other electronic properties of the ZZ and AC edge states are. Via the analytical results we have access to the dependence of the edge-states on AC ribbon width, from which one can obtain how far the edge states penetrate into the bulk of the systems.
This decay length is inversely proportional to $\lambda$, which we obtain numerically by analyzing the energy gap  $\Delta$  at the 1D time-reversal invariant point. This gap  results from  the hybridisation of the edge states localised at opposite edges and its behavior  as a function of the ribbon width $W$ is $\Delta_0~e^{-\lambda W}$ [see the Supplemental material].  
For the case of the armchair ribbon, as discussed above,  
we find that the inverse decay length $\lambda=3 \sqrt{3} t_{SO} / v_F$. As shown in Fig.~\ref{vfermi}d, the numerical results show that $\lambda$ increases linearly with the SOC for AC terminations of both the honeycomb and ruby lattices, in excellent agreement with the analytical calculation (dashed line). 

To obtain analytically the edge state decay length in ZZ ribbons, we solve the full tight-binding Hamiltonian at the 1D time-reversal invariant point $k_x = \pi$ in a semi-infinite ribbon $0<y<\infty$ with open boundary conditions \cite{Liu12}.
Here we use the $k$-dependence of the Hamiltonian \cite{Fu07} given by 
$H({\bf k})=d_0({\bf k}){\mathbb{I}}+\sum_{a=1}^{5}d_a({\bf k})\Gamma^a$, where all $d_a({\bf k})$ are zero except,
$ d_1({\bf k})=t[1+\cos({\bf k}\cdot{\bf a}_1)+\cos({\bf k}\cdot{\bf a}_2)]$, $d_2({\bf k})=t[\sin({\bf k}\cdot{\bf a}_1)+\sin({\bf k}\cdot{\bf a}_2)]$, and 
$d_5({\bf k})=2t_{SO}[\sin({\bf k}\cdot{\bf a}_1)-\sin({\bf k}\cdot{\bf a}_2)-\sin({\bf k}\cdot{\bf a}_1-{\bf k}\cdot{\bf a}_2)]$, 
$\Gamma^1=\sigma^x\otimes\mathbb{I}$, $\Gamma^2=\sigma^y\otimes\mathbb{I}$ and $\Gamma^5=\sigma^z\otimes s^z$.
With the ansatz for the four-component spinorial wave function $\psi = \chi \mathrm{e}^{\lambda y}$, we obtain the secular equation for the zero-energy edge doublet in  ZZ ribbons 
\begin{equation}
{\it Det}  \left|  4 t_{SO} \cosh(\frac {\sqrt{3}}{2} \lambda)  \times \Gamma^5 + t  \times \Gamma^1 \right|  \equiv 0
\end{equation}

As a result, the inverse decay length at ZZ edges depends on the SOC as $\lambda$=$\frac {2}{\sqrt{3}} \cosh^{-1} (\frac {it}{4t_{SO}})$. It is quite remarkable that where the decay length of the edge states at the AC edge is {\it inversely proportional} to the SOC, at the ZZ edge it is {\it proportional} to it. The analytical expression is compared to the numerical results for the honeycomb lattice in Fig.~\ref{vfermi}(c), showing excellent agreement. 
The calculated decay length for the ruby ribbon at 4/6 filling follows a very similar trend. Remarkably the ruby ribbon at 1/6 filling shows almost no dependence on the SOC. 

{\it Conclusions} -- 
Using a combined analytical and numerical approach we have shown that zigzag and armchair edges of honeycomb and ruby QSH nets carry fundamentally different topological edge-states:  the dispersion, velocity and decay length of the edge-states, and their dependence on the spin-orbit coupling strength differ. For the ruby net there is in addition a combined filling {\it and} edge-termination dependence. Particularly the termination dependent decay length that we have established here theoretically provides an interesting and testable prediction for Bi$_{14}$Rh$_{3}$I$_9$ \cite{Rasche13}. In this material the topological edge-states are in principle directly accessible by Scanning Tunneling Microscopy at surface step-edges, as the spin-orbit gap in the QSH layers of this TI material is quite substantial.

The authors thanks M. Richter, M. Ruck, and J.W.F. Venderbos for helpful discussions.


\begin{thebibliography}{}
\bibitem{footnote1} 
Note that for the ruby lattice, due to the larger number of sites per cell (three times larger than for the honeycomb) one needs to divide $N$ by 3 to obtain the width $W$.
\end{thebibliography}
\end{document}